\documentclass[runningheads]{llncs}

\usepackage{eccv}

\usepackage{eccvabbrv}
\usepackage{todonotes}
\usepackage{algpseudocode}
\usepackage{algorithm}
\usepackage{graphicx}
\usepackage{booktabs}
\usepackage[accsupp]{axessibility}  

\usepackage{hyperref}

\usepackage{orcidlink}

\begin{document}

\title{What Makes a Face Look Like a Hat: Decoupling Low-level and High-level Visual Properties with Image Triplets} 

\titlerunning{Decoupling low-level and high-level Visual Properties with Image Triplets}

\author{Maytus Piriyajitakonkij\inst{1,2}\orcidlink{0000-0002-7610-8953} \and
Sirawaj Itthipuripat\inst{3}\orcidlink{0000-0001-9302-0964} \and
Ian Ballard\inst{\dagger*,4}\orcidlink{0000-0003-1814-3141} \and
Ioannis Pappas\inst{*,5}\orcidlink{0000-0002-0168-7014}}

\authorrunning{M.~Piriyajitakonkij et al.}

\institute{Department of Computer Science, The University of Manchester, UK \and
Institute for Infocomm Research (I2R), Agency for Science, Technology and Research (A*STAR), Singapore \and
Neuroscience Center for Research and Innovation (NX), Learning Institute and Big Data Experience Center (BX), King Mongkut’s University of Technology Thonburi, Thailand \and
Department of Psychology, University of California, Riverside, USA \and
Laboratory of Neuro Imaging, University of Southern California, USA
}

\renewcommand{\thefootnote}{} 
\footnotetext[1]{* Senior authors}
\footnotetext[2]{$\dagger$ Corresponding author: ianb@ucr.edu}
\footnotetext[3]{Code: \hyperref[https://github.com/maytusp/triplet_search]{https://github.com/maytusp/triplet\_search}}

\maketitle

\begin{abstract}
\vspace{-0.4cm}
In visual decision making, high-level features, such as object categories, have a strong influence on choice. However, the impact of low-level features on behavior is less understood partly due to the high correlation between high- and low-level features in the stimuli presented (e.g., objects of the same category are more likely to share low-level features). To disentangle these effects, we propose a method that de-correlates low- and high-level visual properties in a novel set of stimuli. Our method uses two Convolutional Neural Networks (CNNs) as candidate models of the ventral visual stream: the CORnet-S that has high neural predictivity in high-level, ``IT-like'' responses and the VGG-16 that has high neural predictivity in low-level responses. Triplets (root, image1, image2) of stimuli are parametrized by the level of low- and high-level similarity of images extracted from the different layers. These stimuli are then used in a decision-making task where participants are tasked to choose the most similar-to-the-root image. We found that different networks show differing abilities to predict the effects of low-versus-high-level similarity: while CORnet-S outperforms VGG-16 in explaining human choices based on high-level similarity, VGG-16 outperforms CORnet-S in explaining human choices based on low-level similarity. Using Brain-Score, we observed that the behavioral prediction abilities of different layers of these networks qualitatively corresponded to their ability to explain neural activity at different levels of the visual hierarchy.  In summary, our algorithm for stimulus set generation enables the study of how different representations in the visual stream affect high-level cognitive behaviors. 
\vspace{-0.4cm}

\vspace{-0.5cm}
\end{abstract}

\begin{figure}[t]
    \centering
    \includegraphics[width=0.62\textwidth]{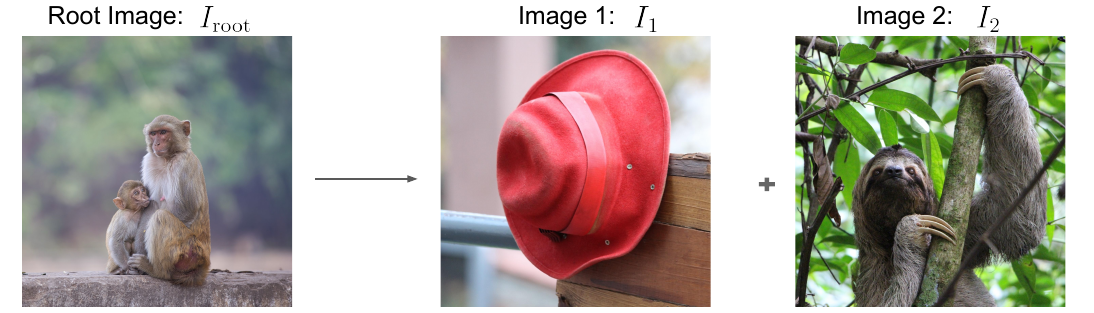}
    \vspace{-0.2cm}
    \caption{Task Description: On each trial, subjects first made a semantic judgment about a root image, i.e., indoor or outdoor. Shortly afterwards, subjects were asked to indicate which of two images they thought was more similar to the root image.}
    \vspace{-0.65cm}
    \label{fig:task}
\end{figure}
\section{Introduction}
\label{sec:intro}
\vspace{-0.2cm}
Low-level visual properties, such as shape, contour, and texture, are integral to visual decision-making. They act as inputs to high-level visual processing goals like object identification and also directly influence preferences and decisions. For instance, a bakery customer might choose a generally less preferred dessert because of its appealing color or texture. However, our understanding of how these lower-level visual properties impact high-level behavior is incomplete.  A better understanding of how lower-level visual information impacts high-level behavior is needed to understand the cognitive and neural bases of visual processing \cite{dicarlo2012brainSolve}.

A key difference between deep neural network models for visual categorization  and the human brain’s visual system is how they make decisions based on abstract representations or features. Modern vision models, such as CNNs \cite{simonyan2014very, kubilius2019cornet-s}, Transformers \cite{dosovitskiy2020vit}, and Recurrent Neural Networks (RNNs) \cite{alkin2024vision}, base their decisions solely on information from the preceding, highest processing layer. In contrast, there are direct connections in humans from the earliest levels of neuronal visual processing to association cortex that influence behavior \cite{trapp2015prediction}. Understanding how the brain integrates low-level and high-level information can inspire researchers to build more brain-like computer vision algorithms \cite{lecun2022path, zador2022toward}.

The ventral visual stream, the object-identification pathway in the brain, is organized hierarchically and neurons in this pathway show response profiles that are strikingly well-modeled with CNNs trained to perform object identification \cite{yamins2014performance, kubilius2018cornet1, kubilius2019cornet-s, schrimpf2018brain}. The Inferior Temporal (IT) cortex is the most high-level visual processing area in the ventral visual stream and contains subregions specialized for detecting faces, body parts, tools, places and other significant visual categories \cite{Bao2020}. Visual stimuli from different categories elicit dissimilar patterns of activity in the IT cortex \cite{kreiman2021biological}. Earlier visual areas such as V2 and V4 process lower-level visual features and are less understood in terms of how these early regions bias visual decision-making. This gap exists partly because the specific visual features that elicit distinct or similar neural patterns in V2 and V4 are not well-studied \cite{dicarlo2012brainSolve, lopez2009roleV2}. Moreover, objects from the same visual category tend to have similar lower-level visual properties, making it difficult to disentangle their unique contributions to decision-making.


We propose a novel approach to generate sets of visual stimuli that decouple high-level and low-level visual similarity. Our approach leverages computational candidate models of the ventral visual pathway, specifically CNNs, which are among the most advanced models for the ventral visual stream \cite{dapello2020simulating, kubilius2018cornet1, kubilius2019cornet-s, yamins2014performance, cross2021using, piriyajitakonkij2021deep, suhaimi2022representation, han2019variational, zhuang2021unsupervised}. We generate stimulus sets of \textbf{triplets of images} composed of a root image and two response images. We aimed to control the relative high- and low-level similarities between the root image and the two response images. This allows us to decorrelate low- and high-level visual properties in a naturalistic set of images, enabling flexible experimental investigation of the role of each level of visual information on behavior. Moreover, our approach permits the comparison of different model architectures in their ability to explain human choices at different levels of visual processing. Most relevantly to the proposed work, CNNs have been previously used to create pairs of images that are parametrized by their similarity on the high-level features and they used these image's similarities to predict brain function \cite{Wammes2022stim_similar}. This approach did not distinguish between high and low-level visual similarity, a key advance of our approach.

We first present our algorithm for stimulus set generation. Second, we collect human behavioral data on our stimulus set. We found that both CORnet-S \cite{kubilius2019cornet-s} and VGG-16 \cite{simonyan2014very} predict human decisions based on high-level visual information, whereas only VGG-16 can account for the influence of low-level visual information on choices. We conclude by comparing these results to BrainScore assessments of these models’ ability to explain neural recording data.

\vspace{-0.4cm}
\section{Methods}
\vspace{-0.2cm}
\label{sec:method}
Our goal is to create stimulus sets of triplets of images, $T = (I_{\text{root}}, I_{\text{1}}, I_{\text{2}})$. Our framework allows users to manipulate the neural network similarity levels between $I_{\text{1}}$ and $I_{\text{2}}$ relative to the root image $I_{\text{root}}$. One can select the similarity model from state-of-the-art deep neural networks and choose which layer in the neural network to represent a particular brain area. The criterion for layer selection is neural predictivity explained below.
\vspace{-0.3cm}
\subsection{Neural Predictivity}
\label{subsec:brainscore}
Neural Predictivity tells how well the response $X$ in a computational model (e.g. a layer’s response) predicts the neural response $y$ in a brain area (e.g. a single neuron activity in V2) given the model and the brain the same images. We use these metrics in BrainScore to compare how our models performed in human data from our study to published neural data. V4 and IT data is from \cite{majaj2015simple} and V1 and V2 data is from \cite{freeman2013functional}. IT responses are collected with 2,560 grayscale images divided into eight types of objects (like animals, boats, cars, chairs, faces, fruits, planes, and tables). Each type includes eight different objects (for example, the ``face'' type has eight different faces). The images were created by placing 3D object models on natural backgrounds. V2 responses are collected with the 9,000 texture stimuli spanning across 15 texture families \cite{freeman2013functional}.
\vspace{-0.4cm}
\subsection{Neural Network Dissimilarity: $D$}
We define the ``neural network dissimilarity'' for high- and low-level layers between $I_{\text{1}}$ and $I_{\text{2}}$ relative to $I_{\text{root}}$ as \ref{eq:ITsim} and \ref{eq:V2sim}:
\begin{equation}
\label{eq:ITsim}
 D_{\text{high}}(I_{\text{root}}, I_{1}, I_{2}) = C(F_{\text{high}}(I_{\text{root}}), F_{\text{high}}(I_{1})) - C(F_{\text{high}}(I_{\text{root}}), F_{\text{high}}(I_{2}))
\end{equation}
\vspace{-0.65cm}
\begin{equation}
\label{eq:V2sim}
     D_{\text{low}}(I_{\text{root}}, I_{1}, I_{2}) = C(F_{\text{low}}(I_{\text{root}}), F_{\text{low}}(I_{1})) - C(F_{\text{low}}(I_{\text{root}}), F_{\text{low}}(I_{2}))
\end{equation}

where $F_{\text{high}}$ and $F_{\text{low}}$ are the high- and low-level layers of the brain model respectively. $C(\cdot,\cdot)$  is the Pearson product-moment correlation coefficient, measuring linear alignment between the model response to $I_{\text{root}}$ and the model response to $I_{\text{1 or 2}}$. If $I_{1}$ and $I_{\text{root}}$ are similar in the high-level layer, the correlation between higher responses to these images is high. If both correlation terms are high, the dissimilarity $D$ will be low, which means both $I_{1}$ and $I_{2}$ are very similar to $I_{\text{root}}$.
\vspace{-0.4cm}
\subsection{Brain-Model-Guided Stimulus Set}
\vspace{-0.6cm}
\begin{algorithm}
\caption{Stimuli selection for each bin}
    \label{algo:stim_select}
	\begin{algorithmic}[1]
        \State $I^{(i)} \in \mathcal{D}$ is an image from the Things dataset $\mathcal{D}$\cite{hebart2023things} 
        \State Select root image index $r \in \{1,2...,|\mathcal{D}|\}$
        \State Select bin indices $b_{\text{high}}, b_{\text{low}} \in \{1,2...,N_{\text{bin}}\}$ 
        \State $Z^{(i)}_{\text{high}}= F_{\text{high}}(I^{(i)})$, $Z^{(i)}_{\text{low}}= F_{\text{low}}(I^{(i)})$  \Comment{Compute layer responses}
        \State $\mathcal{C}_{\text{high}} = \text{corr}(Z_{\text{high}},Z_{\text{high}})$, $\mathcal{C}_{\text{low}} = \text{corr}(Z_{\text{low}},Z_{\text{low}})$\Comment{Correlation matrices}
        \State $S_{\text{high}} = \text{std}(\mathcal{C}_{\text{high}})$, $S_{\text{low}} = \text{std}(\mathcal{C}_{\text{low}})$ \Comment{Standard Deviations}
        \State $\mathcal{T}(b_{\text{low}}, b_{\text{high}}) = \{\}$ \Comment{Initialize a triplet bin \ref{eq:triplet_bin}}
         \For{$i$ in range $|\mathcal{D}|$} \Comment{Loop over the entire dataset}
         \For{$j$ in range $|\mathcal{D}|$}
         \State Randomly select $D_{\text{low}}$ condition in \ref{eq:D_condition} to be positive or negative
          \If{$D_{\text{high}}(I^{(r)}, I^{(i)}, I^{(j)})$ and $D_{\text{low}}(I^{(r)}, I^{(i)}, I^{(j)})$ satisfy \ref{eq:D_condition}}
          \State $\mathcal{T}(b_{\text{low}}, b_{\text{high}}) \cup (I^{(r)}, I^{(i)}, I^{(j)}) $
           \EndIf
         \EndFor
         \EndFor
        \end{algorithmic}
\hspace*{\algorithmicindent} \textbf{Output} $\mathcal{T}(\cdot,\cdot)$
\end{algorithm} 
\vspace{-0.4cm}
We create a stimulus set corresponding to the high-level and low-level layers of the brain model, \ref{eq:ITsim} and \ref{eq:V2sim} respectively. Firstly, we create a triplet container $\mathcal{T}(\cdot,\cdot)$, a function that returns a triplet of stimuli corresponding to given neural network dissimilarity, defined as follows:
\begin{equation}
\label{eq:triplet_bin}
    \mathcal{T}(b_{\text{low}},b_{\text{high}}) = (I_{\text{root}}, I_{1}, I_{2})
\end{equation}
where $b_{\text{low}}, b_{\text{high}} \in \{1,2,...,N_{\text{bin}}\}$ are container bin indices and indicate dissimilarity levels of the sampled triplet $(I_{\text{root}}, I_{1}, I_{2})$. The triplet $(I_{\text{root}}, I_{1}, I_{2})$ is sampled from M triplets in a container bin $b_{\text{low}}, b_{\text{high}}$. Each bin has neural network dissimilarity $D_{\text{low}}$ and $D_{\text{high}}$ as the following conditions
\begin{equation}
\label{eq:D_condition}
D_{\text{low}} = \pm b_{\text{low}}S_{\text{low}} \pm \epsilon ,\;
|D_{\text{high}}| = b_{\text{high}}S_{\text{high}} \pm \epsilon
\end{equation}
where $2\epsilon$ is the size of each bin and $S$ determines the distance between bins. We observe that when $D_{\text{high}}$. is positive, there is a much higher chance that $D_{\text{low}}$ is positive rather than negative, resulting in the high correlation between them. We want to decorrelate $D_{\text{high}}$ and $D_{\text{low}}$. Therefore, the right-hand side of the $D_{\text{low}}$ condition can be either positive or negative with a $50\%$ chance. 
Algorithm \ref{algo:stim_select} describes how stimuli are selected. We exclude selected image indices $(r,i,j)$ in the bins to make sure each image is used only one time in a triplet container. The study design and methods were approved by and followed the ethical procedures of the University of California, Berkeley Committee for the Protection of Human Subjects.
\begin{figure}[t]
    \centering
    \includegraphics[width=0.8\textwidth]{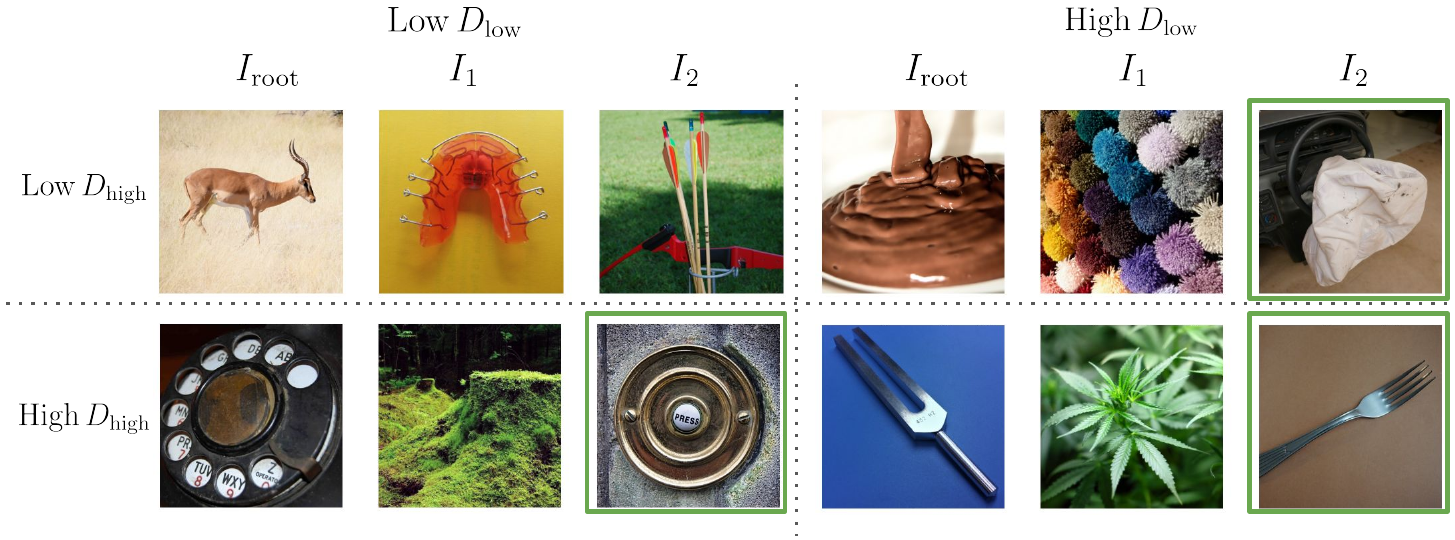}
    \vspace{-0.2cm}
    \caption{Triplet Examples selected from the CC0-Things  \cite{hebart2023things} dataset: (Top-left) $I_{1}$ and $I_{2}$ have the same level of both low- and high-level similarity to $I_{\text{root}}$. (Top-right) $I_{2}$ has higher low-level similarity than $I_{1}$ but the same high-level similarity to $I_{\text{root}}$ ($I_{2}$ is more low-level similar to $I_{\text{root}}$). (Bottom-left) $I_{2}$ has higher high-level similarity than $I_{1}$ but the same low-level similarity to $I_{\text{root}}$. (Bottom-right) $I_{2}$ has both higher low- and high-level similarity than $I_{1}$ to $I_{\text{root}}$.}
    \vspace{-0.5cm}
    \label{fig:triplet_example}
\end{figure}
\vspace{-0.3cm}
\section{Experiment: Dissociating high-level and low-level visual influences on choice}
\vspace{-0.3cm}
We tested whether our stimulus set was able to distinguish behavioral data based on both low-level layers (e.g., V2 area) and high-level layers (e.g., IT area). Participants were tasked to select which image from $I_{1}$ and $I_{2}$ was more similar to the root image $I_{\text{root}}$ as shown in Fig. \ref{fig:task}. Image triplets can vary in their relative low and high level similarity scores, with four extreme cases as shown in Fig. \ref{fig:triplet_example}.

\textbf{Experimental Procedures:} Each trial began with the root image $I_{\text{root}}$ for 2,000 ms, and subjects were instructed to respond based on whether the image was indoor or outdoor. This was meant to encourage encoding of the image and these responses were not analyzed. After a 750 ms inter-stimulus interval, a pair of images appeared. The response image location assignments to the left or right side of the screen were randomized across trials. Subjects ($n = 17$) were instructed to select which image was more similar to the root image. They had 750 ms to respond. If subjects failed to respond to either the root image or the similarity judgment, text was shown instructing subjects to respond more quickly. There were 300 total trials, with a 2,000 ms inter-trial interval. Subjects were instructed that an algorithm determined which image was in fact more similar and that 10 trials would be selected at random and they would be paid \$0.50 for each question they answered correctly according to the algorithm. They were given no feedback to allow them to learn what information the algorithm used; rather, the instructions were meant to motivate subjects. Subjects were paid according to which image had the largest high-level similarity to the root image.

\textbf{CORnet-S:} CORnet-S is designed to replicate the primate ventral visual stream. It achieves high ImageNet classification accuracy compared to other models of similar size and incorporates feedback connections to represent more faithfully the architecture of the ventral visual stream. CORnet-S consists of four blocks, each corresponding to a different area in the ventral pathway: V1, V2, V4, and IT. 

\textbf{VGG-16:} is a CNN architecture composed of 16 layers. It has been widely used for image classification tasks and serves as a benchmark in the field of deep learning. Its operations from the lowest to the highest layers are as follows: Conv1, Conv2, Pool1, Conv3, Conv4, Pool2, Conv5, Conv6, Conv7, Pool3, Conv8, Conv9, Conv10, Pool4, Conv11, Conv12, Conv13, Pool5, FC1, FC2, and FC3. Conv is a convolutional layer, Pool is a max pooling layer, and FC is a fully connected layer.

\textbf{Stimulus Set:} Our stimulus set is created by using the FC3 and Pool3 layers of VGG-16 model. We call it VGGSet. FC3 is the high-level layer and Pool3 is the low-level layer. The key metric is the relative similarity between the two response images and the root: $D_{\text{low/high}}(I_{\text{root}}, I_{1}, I_{2})$. This will give the relative weight of evidence for selecting the left relative to the right image as being more similar to the root. The correlation between this quantity as calculated from FC3 and Pool3 is 0.13. Our stimulus selection algorithm is effective at reducing the correlation between high- and low-level visual properties. Stimuli come from the subset of the Things dataset \cite{hebart2023things} as it is widely used in neuroscience and has rich behavioral and neural data shared among researchers.

\textbf{Analysis}: We analyzed our data using mixed generalized linear models with random-intercepts. We modeled the choice of left versus right image as a binary dependent variable. As independent variables, we included the similarities between the root and the left, relative to right, image derived from both high-level and low-level layers of the model. Including both levels in the same model allows us to test for independent influences of each level of visual information on choice. We also included interaction terms between low-level and high-level similarity. When comparing VGG and Cornet-S, we included model type as a categorical variable as well as its interactions with low-level and high-level similarity.

\vspace{-0.3cm}
\section{Results}
\label{sec:results}
\begin{figure}[t]
    \centering
    \includegraphics[width=0.8\textwidth]{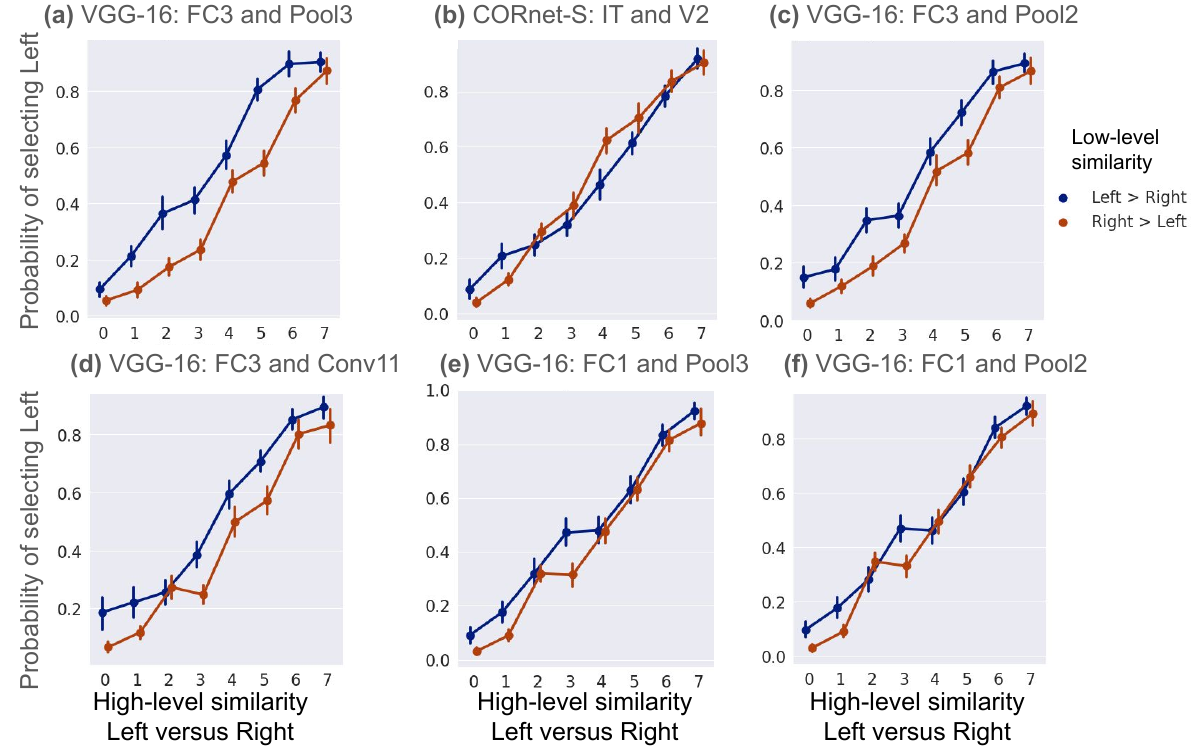}
    \vspace{-0.3cm}
    \caption{Behavioral Results: The y-axis represents the probability that a participant selects the left image $I_{1}$. The x-axis represents the high-level similarity of the left image $I_{1}$ versus the right image $I_{2}$. Mathematically, it is the discretized dissimilarity score $D_{\text{high}}(I_{\text{root}}, I_{1}, I_{2})$. 0 represents the lowest and negative value and 7 represents the highest and positive value, see \ref{eq:ITsim} for the definition of $D$. \textcolor{blue}{Left > Right} refers to the condition $D_{\text{low}}(I_{\text{root}}, I_{1}, I_{2}) > 0$. \textcolor{orange}{Right > Left} refers to the condition $D_{\text{low}}(I_{\text{root}}, I_{1}, I_{2}) < 0$. The pattern of each title is \textbf{network name}: \textbf{high-level layer} and \textbf{low-level layer}.}
    \vspace{-0.2cm}
    \label{fig:behav_result}
\end{figure}

\vspace{-0.2cm}
\textbf{VGG-16 versus CORnet-S:} We found that VGG-16 outperforms CORnet-S in explaining human choices based on low-level similarity, with a significant interaction between model type and both low-level similarity, $Z = 7.8, p=8 \times 10^{-15}$, and CORnet-S outperforms VGG-16 based on high-level similarity, $Z = -23.4, p=2 \times 10^{-16}$, Fig. \ref{fig:behav_result} (a). Interestingly, these behavioral results correspond with neural predicticity scores of these layers, with CORnet-S IT outperforming VGG-16 FC3 in IT neural predictivity, and VGG-16 Pool 3 outperforming CoRnet-S V2 neural predicticity, Fig. \ref{fig:brainscores}. The human subjects are more likely to select the left image when the high-level similarity between the left, relative to right, image and the root image is higher, i.e., $D_{\text{high}}(I_{\text{root}}, I_{1}, I_{2})>0$, $Z = 34.2$, $p = 2 \times 10^{-16}$. Additionally, they select the left image more often if the low-level similarity between the left, relative to right, image and the root image is higher, i.e., $D_{\text{low}}(I_{\text{root}}, I_{1}, I_{2})>0$, $Z = 11.0$, $p = 2 \times 10^{-16}$. This result is depicted in Fig. \ref{fig:behav_result} (a) by showing the categorical preference of whether low-level similarity with the root image was higher for the left, relative to the right image. However, we note that our statistical model used continuous similarity values. Fig. \ref{fig:behav_result} (a) shows that low-level similarity exerts an additive effect: subjects are even more likely to select the left image if it is more similar to the root at both high- and low levels. In contrast, subjects are less likely to select the image with higher high-level similarity if the low-level similarity favors the alternative option, as in Fig. \ref{fig:task}. In contrast, similarities derived from CORnet-S do not show the same pattern as shown in Fig. \ref{fig:behav_result} (b). Whereas high-level similarity metrics derived from CORnet-S do explain choices, $Z = 28.9$, $p = 2 \times 10^{-16}$, low-level similarity metrics do not, $p > 0.1 $.

\textbf{Other layers of VGG-16 also explain choices:} The results are shown in Fig. \ref{fig:behav_result} (c-f). As VGG-16 has many layers, we can select other high-level and low-level layers based on their hierarchy and neural predictivity scores. Therefore, we recomputed  $D_{\text{high}}$ and $D_{\text{low}}$ using different VGG-16 layers. Interestingly, some of the layer combinations exhibit the same behavior only for some high-level similarity levels ($D_{\text{high}}$) unlike the initial VGG-16 layers that we chose to create VGGset. For the high-level layer FC3 and the low-level layer Pool2 (Fig. \ref{fig:behav_result} (c)), the human subjects tend to select the left image when the high-level similarity between the left, relative to right, image and the root image is higher, $Z = 33.8$, $p = 2 \times 10^{-16}$. They also select the left image more often if the low-level similarity between the left, relative to right, image and the root image is higher, $Z = 7.6$, $p = 3.1 \times 10^{-14}$. The same trends apply for other layer combinations with different statistically significant levels as follows: The high- and low-level layer FC1 and Conv11 have $Z = 28.4$, $p = 2 \times 10^{-16}$ and $Z = 9.8$, $p = 2 \times 10^{-16}$ for high- and low-level similarity respectively. The high- and low-level layer FC1 and Pool3 have $Z = 29.7$, $p = 2 \times 10^{-16}$ and $Z = 2.8$, $p = 0.005$ for the high- and low-level similarity respectively. Moreover, the FC1 and Pool2 pairs show the similar statistical values, with $Z = 29.9$, $p = 2 \times 10^{-16}$ and $Z = 2.8$, $p = 0.006$ for high- and low-level similarity trends respectively. Comparing different VGG-16 layers, FC1 outperforms FC3 in explaining high-level similarity's influence on choices, $Z=-20.0, p = 2 \times 10^{-16}$, which is aligned with the high-level neural predictivity score of the IT area that is higher for FC1 than FC3. Pool2 exhibits the same low-level influence on choices as Pool3, $p>0.05$. Conv11 also has the same low-level influence on choices as Pool3, $p>0.1$.


\vspace{-0.4cm}
\section{Discussion}
\begin{figure}[t]
    \centering
    \includegraphics[width=0.4\textwidth]{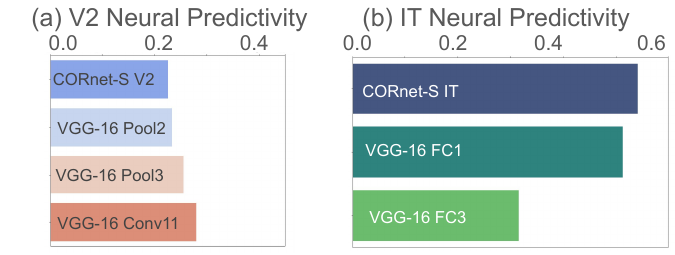}
    \vspace{-0.2cm}
    \caption{Neural Predictivity Score of each model and layer for V2 and IT areas. See \ref{subsec:brainscore} for details.}
    \vspace{-0.5cm}
    \label{fig:brainscores}
\end{figure}

\vspace{-0.2cm}
We found that both the high-level and low-level layers of VGG-16 can account for variance in  human decision making. Overall, we found that VGG-16 explains human decision-making in our task better than CORnet-S, despite CORnet-S having a higher overall neural predictivity score according to Brain-Score \cite{schrimpf2018brain}. We note that although CORnet-S outperforms VGG-16 at explaining IT neural data as shown in Fig. \ref{fig:brainscores}, the VGG's pool3 layer outperforms CORnet-S’s V2 layer at explaining V2 neural data, Fig. \ref{fig:brainscores} (a). We speculate that an increased ability of VGG-16 to explain neural responses in lower-level regions like V2 may account for its ability to explain the influence of low-level visual features on human decisions. Because we did not collect neural data, future research probing the relationship between CNN predictions, neural responses, and human behavior at lower-levels of the visual hierarchy are needed. A weakness of our approach is that our stimulus set was designed using VGG-16 similarities, but not CORnet-S similarities. Additionally, we did not examine every layer in either network. Nonetheless, several lower-level layers in CORnet-S were unable to account for an influence of low-level visual properties on behavior, whereas multiple low-level layers in VGG-16 were able to predict behavior.

The key advance of our approach is that it reduces the correlation between high- and low-level visual features in natural images. Our approach contributes to the growing use of neural networks to generate image sets that permit new ways of addressing questions in neuroscience \cite{Wammes2022stim_similar}. 


\clearpage  

%
%

\bibliographystyle{splncs04}
\bibliography{main}

\end{document}